\documentclass[nogrid]{MBE-onecol}%
\usepackage{algorithmicx}
\usepackage[noend]{algpseudocode}
\usepackage{algorithm}
\usepackage[export]{adjustbox}
\usepackage{titlesec}
\usepackage{listings}
\algrenewcommand\algorithmicindent{.8em}%

\usepackage{textpos}

\newtheoremstyle{exampstyle}
  {.3em} 
  {.3em} 
  {\em} 
  {} 
  {\bfseries} 
  {} 
  {.5em} 
  {} 
  
\theoremstyle{exampstyle} 

\newcommand{\refdef}[2]{\hyperref[#2]{#1}}

\makeatletter
\renewcommand{\paragraph}{%
  \@startsection{paragraph}{4}%
  {\z@}{.8ex \@plus .7ex \@minus .2ex}{-1em}%
  {\normalfont\normalsize\bfseries}%
}
\makeatother

\makeatletter
\let\reftagform@=\tagform@
\def\tagform@#1{\maketag@@@{(#1\unskip\@@italiccorr)}}
\renewcommand{\eqref}[1]{\textup{\reftagform@{\ref{#1}}}}
\makeatother

\usepackage{pdfpages}
\usepackage{url}
\jshort{mst}

\setboolean{@twoside}{false}
\volname{}

\jvolume{0}

\jvol{}

\jissue{0}

\pubyear{2016}

\access{MBE Advance Access}

\usepackage{textpos}

\begin{document}

\title[Local coalescent-based posterior probability]
{Fast coalescent-based computation of local branch support from 
quartet frequencies}

\author[Sayyari and Mirarab]{Erfan \surname{Sayyari}$^{1}$ and Siavash Mirarab$^{\ast,1}$}

\address{$^{1}$Department of Electrical and Computer Engineering,
University of California at San Diego, 9500 Gilman Drive, La Jolla, CA 92093}

\history{}

\coresp{E-mail: smirarab@ucsd.edu.}

\abstract{
Species tree reconstruction is complicated by effects of Incomplete Lineage Sorting (ILS), commonly modeled by the multi-species coalescent model. While there has been substantial progress in developing methods that estimate a species tree given a collection of gene trees, less attention has been paid to fast and accurate methods of quantifying support. In this paper, we propose a fast algorithm to compute quartet-based support for each branch of a given species tree with regard to a given set of gene trees. We then show how the quartet support can be used in the context of the multi-species coalescent model to compute i) the local posterior probability that the branch is in the species tree and ii) the length of the branch in coalescent units. We evaluate the precision and recall of the local posterior probability on a wide set of simulated and biological datasets, and show that it has very high precision and improved recall compared to multi-locus bootstrapping. The estimated branch lengths are highly accurate when gene tree estimation error is low, but are underestimated when gene tree estimation error increases. Computation of both the branch length and the local posterior probability is implemented as new features in ASTRAL.
}


\maketitle

\newcommand{\LS}{\mathcal{L}} 
\newcommand{\GT}{\mathcal{G}} 
\newcommand{\qp}{Q} 
\newcommand{\tr}{{R}} 
\newcommand{\TR}{T} 
\newcommand{\ql}{q} 
\newcommand{\qt}{t} 
\newcommand{\fq}{n} 
\newcommand{\FQ}{\bar{n}} 
\newcommand{\br}{\qp} 
\newcommand{\pr}{P} 
\newcommand{\qs}[1]{\psi(#1)} 
\newcommand{\qsr}[1]{\psi(#1)} 
\newcommand{\third}{{\frac{1}{3}}} 
\newcommand{\gns}{n}
\newcommand{\sps}{l}
\newcommand{\FQV}{\bar{N}}
\newcommand{\fqs}[1]{
\ifx&#1&%
\bar{z}%
\else%
z_{#1}%
\fi%
}
\newcommand{\FQS}{\bar{Z}}

\begin{textblock*}{3cm}(7.7cm,-16.5cm)
  \fbox{\footnotesize \textbf{MBE Advance Access published April 15, 2016}}
\end{textblock*}
\section{{Introduction}\label{sec:Intro}}
The multi-species coalescent model (MSC) of \citet{Rannala2003} has emerged as 
the standard  method used for reconstructing  the species trees in the presence of gene tree discordance due to Incomplete Lineage Sorting (ILS)~\citep{Maddison1997,Degnan2006}.
Many methods have been developed to estimate species
trees under the MSC \citep[e.g.,][]{Heled2010,SVDquartets,binning,snapp}.
The most scalable family of MSC-based methods are based on a two-step process where gene trees are first estimated independently for each gene and are then combined to build  the species tree using a {\em summary method}. Many of the summary methods are statistically
consistent and thus converge in probability to the true species
tree as the number of input error-free gene trees increases;
examples of consistent methods include
ASTRAL~\citep{astral,astral2}, 
BUCKy-population~\cite{Larget2010}, 
GLASS~\cite{glass},
MP-EST~\citep{mpest},
NJst and ASTRID~\citep{njst,astrid},
and STAR~\cite{star}.
While some methods (e.g., MP-EST) can estimate branch lengths in coalescent units,
others only infer the topology. 
The traditional concatenation
approach (where all genes are put together
in a supermatrix) can produce
high support for incorrect branches~\cite{Roch2014,KubatkoDegnan2007},
and the main goal of statistically consistent summary
methods is to address this shortcoming. 
However,
despite the progress in developing methods
for species tree reconstruction,
little attention has been paid to 
methods of calculating support.

Bayesian methods 
\citep[e.g.,][]{Heled2010,best}
readily provide support but remain computationally challenging.
Calculating support through
bootstrapping \cite{bootstrap},
while still computationally expensive, is not prohibitively
slow and is easily parallelizable. 
\citet{Seo2005} proposed a  
multi-locus bootstrapping (MLBS) procedure that produces bootstrap replicates
by first resampling genes 
and then sites within those sampled genes. 
\citet{Seo2008} later studied the accuracy of the MLBS approach in the
context of distance-based tree reconstruction for 4-taxon trees 
and explored other strategies where only genes or only sites
were resampled. 
These earlier works did not consider ILS as the cause
of discordance; nor did they use summary methods. 
Nevertheless, the community has adopted 
MLBS as a standard way of estimating
support using ILS-based summary methods; 
most biological studies using summary methods rely on site-only or site/gene MLBS
\citep[e.g.,][]{Song2012,1kp-pilot,avian,Prum2015}.

Recently, \citet{mrl-sysbio} studied the reliability of MLBS support values as a measure of accuracy in
simulation studies, and documented both under-estimation and
over-estimation of support (for low and high support branches, respectively)
using MP-EST~\cite{mpest} and supertree methods
such as MRP~\cite{Ragan1992} and MRL~\cite{mrl}. 
\citet{mrl-sysbio} also observed better species tree accuracy
when a summary method was run directly on ML gene trees,
 compared to running the summary method on 
bootstrapped gene trees first and then taking a consensus.
This observation led to the conjecture that MLBS
might give biased estimates of the support. 
Furthermore, 
\citet{WSB} found in simulation studies that 
false positive branches can sometimes have high MLBS support and that many
true branches tend to have low support.
Obtaining low support for true branches
should not be a cause of concern if the lack of support is caused by insufficient data;
however, when low support is caused by
underestimation, we need better methods of 
quantifying support. 

In this paper, we show that 
using properties
of the MSC on four taxa (quartets), we can
derive support values that are more precise and
more
powerful than MLBS, and are much faster to compute.  
Under the MSC, quartet trees do not have anomaly zones~\cite{Allman,Degnan2013},
meaning that the most probable gene tree is 
identical to the species tree for any quartet. 
Exploiting this property,
some summary methods 
break up gene trees into quartet trees. 
For example, 
ASTRAL~\cite{astral2}, a summary method
 used in many recent studies~\citep[e.g.,][]{1kp-pilot,Laumer2015,Giarla2015,Prum2015,Hosner2011,Grover2015,Streicher2015,Andrade2015},
finds the species tree that shares the maximum
number of induced quartet
trees with  gene trees. 

\begin{figure}[t]
\begin{center}
\includegraphics[width=0.47\textwidth]{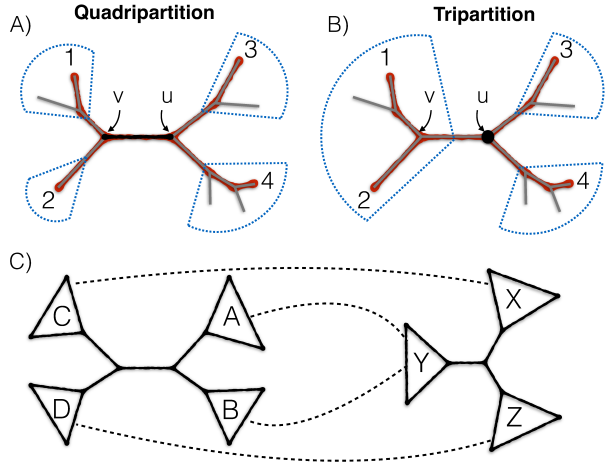}
\end{center}
\caption{Quadripartitions and tripartitions.
A) An internal branch (black, in the middle)
divides the set of leaves into a quadripartition.
A quartet of leaves $\{1,2,3,4\}$ induces
a quartet tree (in red) with 
two internal nodes that map to two nodes
in the larger tree (here,  $u$ and $v$). 
B) An internal node, here $u$, divides leaves
into a tripartition; a selection of two leaves from
one side and one from each remaining side
gives a quartet mapped to that tripartition.
Each quartet tree also maps to a second tripartition ($v$).
C) An example mapping between a quadripartition
(e.g., from the species tree) and a tripartition
(e.g., from a gene tree); 12 such mappings exist. 
Note that by finding all quartets of the form
$(a,b,c,d); a\in A\cap Y, b\in B\cap Y, c\in C\cap X, d\in D\cap Z$,
we can find all quartets around the quadripartition that
are mapped to the tripartition with this mapping.}%
\label{fig:basics}%
\end{figure}

We introduce a new method for computing support for  species 
tree branches with regard to a set of unrooted gene trees by calculating Bayesian posterior probabilities. 
Our support values, which we call  {\em local posterior probabilities},
are computed based on gene tree quartet frequencies. 
For each internal branch of the given species tree, we {\em assume}
that the four {\em sides} of the branch (Fig.~\ref{fig:basics}) are correct,
and therefore three topologies are possible around that branch. 
We introduce a fast algorithm to 
compute a {\em quartet support} for each of those three alternatives
in $\Theta(\gns \sps)$ time (where $\sps$ is the number
of species and $\gns$ is the number of genes). 
We then use the quartet support for each alternative topology
to derive the posterior probability 
that it is the correct species topology. 
 Besides producing measures of support, quartet frequencies  can be used to derive estimates of internal branch lengths 
 in coalescent units.

Our calculation of posterior probabilities is 
analogous to characterizing a biased die.
If we toss a three-faceted biased die
 $\gns$ times, 
our belief in whether the die is biased towards 
 a certain side should
 depend on
 the number of tosses and also on the bias of the die
 (less bias requires more tosses).
 Similarly, a short branch in the species
 tree will result in high discordance,
 and will need many genes to resolve it with
 high confidence. 
 On the other hand,
 considering only the MSC and ignoring issues such as long branch
 attraction, long
 branches can be easily reconstructed
 confidently even with few genes. 

We show using simulated and empirical datasets that 
the local posterior probability 
estimated by our approach is a reliable measure of accuracy. 
We show that very few highly supported branches are incorrect.
Moreover, with a sufficient number of genes, most correct branches have high support. 
Importantly, we test our methods
under conditions where assumptions of 
our model are violated and show that it remains reliable. 
Our method is available as part of ASTRAL (\url{https://github.com/smirarab/ASTRAL/}), 
which now estimates species tree topologies, branch lengths, and 
local posterior probabilities.

\section{New Approaches}

\paragraph{Definitions:}
Throughout this paper, we only consider unrooted trees. 
Let $\LS$ be the set of $\sps$ leaves (i.e., taxa).
Each branch of a tree $\TR$ creates a bipartition on 
$\LS$, and we say that 
each of the two partitions is a {\em cluster} in $\TR$. 
Each {\em internal} branch divides $\LS$ into four clusters,
creating a quadripartition (Fig.~\ref{fig:basics}). 
Similarly, an internal node divides $\LS$ into three clusters, creating
a tripartition (Fig.~\ref{fig:basics}).
Any quartet $\ql$ of taxa induces a quartet tree $\qt$ on $\TR$.
The two internal nodes of $\qt$ correspond to two internal nodes in $\TR$.
When those two internal nodes are the two sides of a single branch in $T$, 
we say that the quartet $\ql$ is {\em around} that branch.
The set of quartets  
around a given quadripartition can be 
built by
enumerating all selections of one leaf from each 
of its four clusters.

\paragraph{Problem statement:}
We are given a set of $\gns$ gene trees evolved on an unknown true binary species tree
according to the MSC model. 
Our aim is to
{\em score} a given internal branch represented 
as a quadripartition $\qp$ to estimate: 
\begin{arabiclist}
\item the probability that $\br$ is in the true species tree,
assuming clusters of $\qp$ are each correct,
\item the length of $\br$ in coalescent units, assuming $\br$ is a correct
branch in the species tree.
\end{arabiclist}

\paragraph{Assumptions:}
We assume evolution is tree-like and true gene trees differ
from the species tree only due to ILS, as modeled by the MSC. 
We also assume we are given
an unbiased sample of true gene trees. 
On real data, we need to instead estimate
gene trees from sequence data, and further, it is not always clear that our sample is unbiased, 
nor that gene trees are generated by the MSC. 

Importantly, we further assume that all four clusters around the 
branch we are scoring are correct. 
This assumption, which we refer to
as the {\em locality assumption}, 
makes our computations tractable for large datasets.  
Similar assumptions have been made in past for 
fast calculation of local support in the context of maximum likelihood (ML) tree reconstruction
from sequence data; e.g., aLRT in PhyML~\cite{Guindon2009} and SH-like support in
FastTree-II~\cite{fasttree-2}. 
Note that to test our method, we use data
that violate the locality assumption 
and we also use estimated
gene trees in addition to true gene trees
generated by the MSC.

\subsection{Calculation of local posterior probability}

\paragraph{Quartet trees:}
Our approach is based on analyzing quartets defined around the branch $\br$.
For any quartet of leaves around $\br$, 
we have three possible topologies, which we will
call $\qt_1$, $\qt_2$, and $\qt_3$. 
In the MSC model, the 
quartet topology found in
the true species tree
has the highest probability of
appearing in gene trees~\cite{Allman}, and the two alternative topologies have identical probabilities.
Furthermore, if the total branch length (in coalescent units) between the two internal nodes of the quartet is $d$, 
the probability of the dominant quartet
topology in gene trees is $\theta=1-\frac{2}{3}e^{-d}>\third$,
and the probabilities of both alternative topologies are $\frac{1-\theta}{2}=\frac{1}{3}e^{-d}<\third$.

We refer to the number of times $\qt_1,\qt_2,$ and $\qt_3$
are induced in gene trees as {\em quartet frequencies},
shown as $\FQ=(\fq_1,\fq_2,\fq_3)$; note $\sum_1^3 \fq_j =\gns$. 
In the MSC model, conditioned on the species tree,
gene trees are independent; 
Thus,
%
$\FQ$ can be modeled as a multinomial random variable $\FQV$, 
with parameters $\theta$, $\frac{1-\theta}{2}$, $\frac{1-\theta}{2}$,
where $\theta$ corresponds to the species tree topology, 
and $\frac{1-\theta}{2}$ to the alternative topologies. 
A similar model is used in the maximum pseudo-likelihood 
approach of \citet{mpest}  for triplets.

\paragraph{Multiple quartets:}
There are $m=\prod_1^4 m_i$ quartets
around branch $\qp$, where $m_i$ is the size of a
cluster of the quadripartition of $\qp$.
Note that
we can rearrange clusters of $\qp$ to 
obtain two alternative quadripartitions, 
which we call $\qp_2$ and $\qp_3$.
Let $\FQ_i=(\fq_{1i},\fq_{2i},\fq_{3i}), 1\le i \le m$ be
quartet frequencies for all $m$ quartets around branch $\br$ such that
$\fq_{1i}$, $\fq_{2i}$, and $\fq_{3i}$  correspond to the topologies $\qp$, $\qp_2$ and $\qp_3$, respectively.

Each $\FQ_i$ can be modeled as a
multinomial random variables $\FQV_i$, and 
$\FQV_i$s are identically distributed.
To use all $\FQ_i$ values, 
one approach is to assume 
they are also all independent, and
model $(\sum\fq_{1i},\sum\fq_{2i},\sum\fq_{3i})$ as 
observations from a multinomial with
$m\times \gns$ trials.
The independence assumption would clearly be incorrect; 
topologies of different
quartets around a branch heavily depend on each other (Fig. S15 shows an example).
Quartets around a branch are dependent 
even when the locality assumptions holds.
A big problem with the independence assumption is that it inflates
confidence because the number of observations (i.e., die tosses)
becomes $m\times \gns$ instead of $\gns$, thereby greatly increasing
posterior probabilities (note $m\ge \sps-3$).
Moreover, the dependence of various quartets on each other
is intricate and hard to model. 

To avoid inflating posterior values by assuming independence,
we take the opposite conservative approach. 
We assume that a hidden 
random variable $\FQS$ 
gives a single vector of ``true'' quartet frequencies around $\qp$
and treat each $\FQV_i$ as a noisy estimate of $\FQS$.
Thus, 
$\FQS$ follows a multinomial
distribution with $\gns$ tries (irrespective of $m$)
and $\FQV_i = \FQS+\bar{Y}$
for $1 \leq i \leq m$  
where $\bar{Y}$ is a noise term with zero expectation.
In the die analogy, 
we assume the die is tossed $\gns$ times,
and for each toss,
we read the outcome $m$ times, each time with some noise.
Ideally, we should have a noise model and 
compute the posterior with respect to the
given
$\FQV_i$ values by marginalizing over $\FQS$. However, a good noise
model is not available and the resulting problem
becomes hard to solve. Instead,
we treat the expected value of $\FQS$ as
an observed value, and empirically estimate it by averaging:
\begin{equation}
\fqs{j}=\frac{\sum_1^m \fq_{ji}}{m} ~~ \text{for}~ j \in \{1,2,3\}
\label{eq:fqs}%
\end{equation}\label{lm:multi}
At the end of this section, we will introduce an efficient $\Theta(\gns \sps)$ algorithm to compute $\fqs{}$.

%
\begin{flLem}
Let $(\theta_1,\theta_2, \theta_3)$
denote  parameters of the true multinomial distribution generating $\FQS$. Note $\sum_1^3 \theta_i=1$ and the
two lower $\theta_i$s are identical, and recall
$\fqs{1}$ corresponds to the topology of $\br$.
\begin{equation}
\pr(\theta_1>\third|\FQS=\fqs{})=\frac{\int_\third^1 \pr(\FQS=\fqs{}|\theta_1=t)f_{\theta_1}(t)\mathrm{d}t}{\pr(\FQS=\fqs{})}
\label{eq:pp}%
\end{equation}
where $f_{\theta}$ is the prior PDF. The likelihood term is:
\begin{equation}
\pr(\FQS=\fqs{}|\theta_1=t)
=\Gamma t^{\fqs{1}}(\frac{1-t}{2})^{\gns-\fqs{1}}
\label{eq:lik}%
\end{equation}
where $\Gamma=\frac{\Gamma(\gns+1)}{\prod_1^3 \Gamma(\fqs{j}+1)}$, and marginal probability is:
\begin{align}
\pr(\FQS={\fqs{}})&= \sum_{j=1}^3 \int_\third^1 \pr(\FQS={\fqs{}}|\theta_j=t)f_{\theta_j}(t)\mathrm{d}t \nonumber\\
&=\Gamma \sum_{j=1}^{3}\int_\third^1 
 t^{\fqs{j}}(\frac{1-t}{2})^{\gns-\fqs{j}}  f_{\theta_j}(t)\mathrm{d}t.
\label{eq:marg}%
\end{align}\label{lm:lpp}
\end{flLem}
The proof is given in supplementary material. 
$\br$ is in the species tree iff
$\theta_1>\third$; thus, with Lemma~\ref{lm:lpp} and a 
prior we can compute the 
posterior probability.

\paragraph{Prior:} 
%
In absence of extra reliable information about the species tree topology,
which is the most common scenario, 
the use of an uninformative prior is justified. 
An uninformative prior would require that 
the three topologies are equally likely
(i.e., $\pr(\theta_1>\third)=\pr(\theta_2>\third)=\pr(\theta_3>\third)=\third$).
%
Based on Theorem~3.3 from \citet{Stadler2012},
we can prove (Supplementary material):
\begin{flLem}
If the species tree is generated using the Yule process
with rate $\lambda$,
branch lengths are exponentially distributed, 
and for $t\geq \third$:
\begin{equation}
f_{\theta_j}(t) = \lambda (3\frac{1-t}{2})^{2\lambda-1} \label{eq:prior}
\end{equation}
\label{lem:prior}  
\end{flLem}
We use
(\ref{eq:prior}) throughout the paper as the prior 
($\lambda=\frac{1}{2}$ 
gives a flat prior). 
Note that we need that branch
lengths in {\em coalescent units} follow
properties of the Yule process; this can be 
achieved if lengths measured by
the number of generations follow the Yule
process and $N_e$ is constant for all branches. 

\paragraph{Local posterior probability:}
We now conclude: 
\begin{flthem}
Given 1) a set of $\gns$ gene trees generated by the MSC on
a model species tree generated by 
the Yule process with rate $\lambda$ and 2) an internal branch
represented by a quadripartition $\qp$ 
where the four clusters around
$\qp$ are each present in the species tree,
let $\fqs{}=(\fqs{1},\fqs{2},\fqs{3})$ be the average quartet frequencies
around $\qp$ (where $\fqs{1}$ corresponds to the topology of  $\qp$); the local posterior probability
that the species tree has the topology given by $\qp$ is:
\begin{equation}
\pr(\qp|\FQS=\fqs{})=\frac{h(\fqs{1})}{h(\fqs{1})+2^{\fqs{2}-\fqs{1}}h(\fqs{2})+2^{\fqs{3}-\fqs{1}}h(\fqs{3})}
\label{eq:ppf}%
\end{equation}
for 
$h(x)=\mathbf{B}(x+1,\gns-x+2\lambda)(1-I_\third(x+1,\gns-x+2\lambda))$.
Here, $\mathbf{B}(\alpha,\beta)$ is the beta function, 
and $I_x$ is the regularized incomplete beta function.
\label{thm:pp}
\end{flthem}

\begin{proof}[\textit{Proof (sketch)}]
With locality assumption,
$\FQS$
follows a multinomial distribution with 
parameters $(\theta_1,\theta_2,\theta_3)$.
Lack of anomaly zones for unrooted quartets, shown
by~\citet{Allman} means that $\qp$ is in the species tree
iff $\theta_1>\third$. Thus, by Lemma~\ref{lm:lpp}
we can use (\ref{eq:pp}), (\ref{eq:lik}), and (\ref{eq:marg}) to compute
the local posterior probability of $\qp$.
By the assumption that (coalescent unit) branch lengths
in the species tree are generated by the Yule process,
and by Lemma~\ref{lem:prior}, our prior becomes
the equation shown in (\ref{eq:prior}). 
Calculation of (\ref{eq:ppf})
follows from manipulating (\ref{eq:pp}), (\ref{eq:lik}),
and (\ref{eq:marg}), as detailed in the Supplementary material. 
\end{proof}

\begin{figure}[t]
\begin{center}
\includegraphics[width=0.48\textwidth]{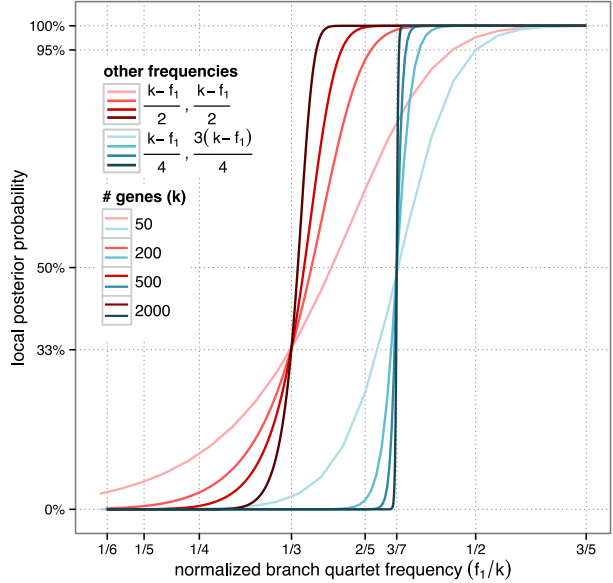}
\end{center}
\caption{The local posterior probability of a branch 
as a function of its
normalized quartet support for varying
numbers of genes.
Red lines:  alternative topologies
have equal frequencies (thus,
conform to properties of the MSC
for x greater than 1/3);
Blue lines: alternative topologies
don't have equal frequencies (contrary to the MSC). }%
\label{fig:example}%
\end{figure}

\paragraph{Examples:}
The local posterior probability (pp)
is a function of the number of genes
and the quartet frequency of a branch. 
As Figure~\ref{fig:example} shows,
a branch that appears in 40\% of gene trees 
has a low 66.1\% pp if 50 gene
trees given (and alternative topologies
are equally frequent); however, with 200 or 500 
genes, the same branch will have 
93.0\% or 99.7\% pp, respectively. 
Thus, a high discordance branch
 where 60\% of genes
do not agree with the species tree
can still be resolved with high confidence 
given enough genes. 
Moreover, the posterior probability is affected
not just by the frequency of the topology being
scored, but also by the frequency of the two alternatives. 
For example, if our branch of interest
appears in 40\% of gene trees, but a second
alternative appears in three-quarters of the
remaining genes (i.e., 45\% of genes),
the branch with 40\% frequency 
has only  a 1.90\% pp (Fig.~\ref{fig:example}).

\subsection{Calculation of branch length}

Given the true parameter $\theta$ for a correct branch, its
length in coalescent units~\citep{Degnan2006}
is simply $-\ln \frac{3}{2}(1 - \theta)$. 
Thus, we can prove (see Supplementary material):

\begin{flthem}
Under conditions of Theorem~\ref{thm:pp}, and 
assuming the branch represented by $\qp$
is in the species tree, 
the ML estimate for its length is
$-\ln \frac{3}{2}(1 - \frac{\fqs{1}}{\gns})$ and the MAP
estimate is $-\ln \frac{3}{2}(1 - \frac{\fqs{1}}{\gns+2\lambda})$
when $\fqs{1}\ge \frac{\gns}{3}$ and $\fqs{1}\ge \frac{\gns+2\lambda}{3}$, respectively;
otherwise,  ML$=$MAP$=0$.
\end{flthem}

\subsection{Calculation of quartet support}
We now discuss how $\fqs{}$ defined in (\ref{eq:fqs}) can be efficiently computed. 
Note that in the worst case, 
there can be $\Theta(\sps^4)$ quartets around a single
branch. Thus, simply enumerating all $\fq_{ij}$
values and then getting the average can be very slow. 

As noted before, each quartet 
around $\qp$ 
has each of its four leaves
drawn from a different cluster of $\qp$.
Recall also that internal nodes of a tree
produce a tripartition, and as \citet{astral} 
pointed out, any selection of two leaves from one side
of the tripartition and one leaf from each remaining side 
gives a quartet tree mapped to that tripartition
(Fig.~\ref{fig:basics}).
Let $\qs{\qp}$ and $\qsr{\tr}$ give the set of quartet trees around
a quadripartition and mapped to a tripartition, respectively. 
Any quartet tree around $\qp$ that is induced by a
gene tree will be mapped to two internal nodes in
that gene tree (Fig.~\ref{fig:basics}). Thus, 
\begin{equation}
\fqs{1} = \frac{1}{2m}\sum_{g=1}^{\gns} \sum_{u=1}^{\sps-2}
|\qsr{\tr_u^g}\cap \qs{\qp}| \label{eq:fqs1}
\end{equation}
where $\tr_u^g$ is a tripartition for node 
$u$ in the gene tree $g$, and $m$ is the number
of quartets around $\qp$.  

We can efficiently compute the number of quartet topologies
around $\qp$ that appear also in 
a tripartition $\tr$ (i.e., $|\qsr{\tr_u^g}\cap \qs{\qp}|$)
without computing $\qs{.}$ sets. 
We define a mapping between clusters 
around $\qp$ to clusters
in $\tr$: map two sister clusters in $\qp$ 
to a cluster in $\tr$, and map the remaining two 
clusters in $\qp$ to the remaining two clusters
 in $\tr$ (Fig.~\ref{fig:basics}). 
 For example, let $\qp=AB|CD$ and $\tr=X|Y|Z$;
 a possible matching is to map $A$ 
and $B$ to $Y$, $C$ to $X$, and $D$ to $Z$. 
There are 12 such matchings between $\qp$ and $\tr$.
For each matching, 
we can compute the number of quartet
trees around $\qp$ that appear in $\tr$ by multiplying
the sizes of the intersection of pairs of clusters
in $\qp$ and $\tr$ that are mapped to each other. 
By enumerating all 12 matching
and summing the resulting numbers, 
we get $|\qsr{\tr_u^g}\cap \qs{\qp}|$.
Since finding the intersection of two clusters
requires $\Theta(\sps)$, computing (\ref{eq:fqs1}) would require 
$O(\sps^2\gns)$ running time. 
However, we can do better. 
\citet{astral2} have introduced a $\Theta(\gns \sps)$ 
algorithm to compute
a sum similar to (\ref{eq:fqs1}) for scoring a tripartition,
based on a postorder traversal of gene trees (instead
of analyzing each $\tr_u^g$ separately).  
This algorithm can be  adopted
here to compute (\ref{eq:fqs1}) in $\Theta(\gns \sps)$ (see Fig. S14 for the algorithm).
Since scoring a tree requires scoring $\sps-3$ branches, 
\begin{flthem}
Computing branch lengths and local posterior
probabilities of a tree requires
$\Theta(\sps^2\gns)$ time. 
\end{flthem}

\subsection{Other considerations}

We implemented our methods in 
ASTRAL,
using the Colt~\cite{colt} package for numerical 
computations.
Handling missing data and unresolved gene trees
requires extra care. 
When gene trees have missing data, 
to compute (\ref{eq:fqs1}), instead of setting $m$
to the number of quartets around $\qp$, we need
to set it to the average number of quartets
present in gene trees (i.e., $\frac{1}{\gns}\sum_1^3 \fqs{j}$).
Moreover, missing data can cause some genes to miss
{\em all} quartets around $\qp$; to account
for this, we allow a different $\gns$ for each branch, 
and set it to the number of genes that include at least
one of the quartets around $\qp$. 
To handle unresolved gene trees,
similar to ASTRAL-II,
we need to score the quadripartition against all ${d \choose 3}$
tripartitions around a polytomy with degree $d$.


\section{Materials and Methods}

\subsection{Datasets}

We use both simulated and biological datasets. 

\subsubsection{Simulated data}
We use two sets of simulated datasets from previous publications:
the 200-taxon dataset (called A-200 here) from~\citet{astral2} and an avian dataset with 48 taxa
from~\citet{binning}.
A-200 enables us to test accuracy under heterogeneous
conditions with many species, and the avian dataset is used
to compare local posterior against MLBS. 
For both datasets, gene trees are simulated using the MSC, and their branch lengths are then
adjusted to be in substitution units and
to deviate from the strict molecular clock.
Sequence data are next simulated on the
modified gene trees using GTR+$\Gamma$,
and ML gene trees are estimated from the data. 
On the avian dataset, 
bootstrapped gene trees are also available.
For both datasets, in addition
to true species trees, 
we have estimated species trees 
(ASTRAL and NJst on
estimated gene trees, and 
concatenation using ML). 
We show results for ASTRAL and true species tree here
and show the rest in the supplementary material.

\begin{table}[!t]
\tableparts{\caption{Properties of simulated
datasets. }\label{tab:datasets}}
{\begin{tabular*}{\columnwidth}{@{\extracolsep{\fill}}lld{1,0}d{1,0}d{1,0}d{2,0}d{2,0}d{8,0}@{}}\toprule 
&Cond.&$\sps$ &$\gns$&R&ILS&GE& SE   \\
\colrule 
A-200& Low-ILS&201&1000 &100 & 15\%& 25\%&6\%,4\%,3\%\\
[0.1pt]
A-200& Med-ILS&201&1000 &100 &34\%& 31\% &9\%,6\%,4\%\\
A-200& High-ILS&201&1000 &100 & 69\%& 47\%&19\%,10\%,6\% \\
Avian & 1500bp&48&1000 &20& 47\% &31\% &5\%\\
Avian & 1000bp&48&1000&20& 47\% & 39\% &6\%\\
Avian & 500bp & 48&1000&20&  47\% & 54\%&8\%\\
Avian & 250bp &48&1000&20& 47\% &67\% &15\%\\
\botrule
\end{tabular*}}
{$\sps$: number of species; $\gns$: maximum number of genes
(A-200 also has 50 and 200);
$R$: number of replicates;
$ILS$: average normalized RF~\cite{RF} distance (AD) between 
the true species tree and true gene trees;
$GE$: AD between true and estimated gene trees;
$SE$: AD between true and ASTRAL species trees
(for A-200, with 50, 200, and 1000 genes, respectively).}
\end{table}

\paragraph{A-200}:
The 201-taxon datasets (200 ingroups plus an outgroup, 
treated like other taxa here) 
are simulated using SimPhy~\cite{simphy}, and
has three levels of ILS (Table~\ref{tab:datasets}),
with true discordance that
ranges from very low to very high (Fig. S1).
Each replicate of the simulation has its own species tree, 
and the ILS level is controlled by changing
the tree length (500k, 2M, and 10M generations). \citet{astral2} generated species trees using the
Yule process with two speciation rates ($10^{-6}$
and $10^{-7}$ per generation)
for each tree length, but here we combine
the two rates into one dataset to get twice the number of replicates.
SimPhy automatically introduces deviation from the
strict clock by
drawing species, gene, and gene/species-specific rate multipliers
from predefined distributions. Similarly,  the
number of sites for each gene is randomly chosen.
See \citet{astral2} for  details.
ML Gene trees are estimated using FastTree-II~\cite{fasttree-2},
with a wide range of estimation error
(Table~\ref{tab:datasets} and Fig. S1).  
Species trees are estimated 
based on all 1000 genes per replicate,
or on subsets of 200 or 50 genes. 
The ASTRAL species trees error for various datasets ranges between average 
3\% and 19\% 
(Table~\ref{tab:datasets}). 

\paragraph{Avian:} The avian dataset has
48 taxa, and is simulated to emulate the whole-genome dataset of~\citet{avian},
possibly overestimating the true amount of ILS~\cite{binning,Gatesy2014}. 
Here, we use four conditions, 
with 20 replicates that each 
includes 1000 genes, all simulated based on the same avian-like species tree. 
Our four conditions differ in terms of the number of sites
per gene (250bp, 500bp, 
1000bp, or 1500bp), creating varying levels of gene tree estimation error (Table~\ref{tab:datasets}).
ML Gene trees are estimated using RAxML~\cite{Stamatakis2014}, and 200 replicates of bootstrapping are performed.
Average ASTRAL species tree error ranges from
5\% to 15\%, depending on the gene tree error (Table~\ref{tab:datasets}). 
We used site-only MLBS to get BS support values.
A single branch in our true tree was 
extremely short (almost a polytomy, with a length of $10^{-6}$).
When discussing branch length accuracy, we ignore that 
branch; results including that branch will also be shown for
completeness. 

\subsubsection{Biological dataset:}
We reanalyze 
four published datasets:
a 103-taxon 424 gene plant dataset by 
\citep{1kp-pilot}, 
a 46-taxon 310 gene angiosperm dataset
by \citet{xi},
a 48-taxon 2022 (binned) supergene tree dataset
by \citet{avian}, and
a 201-taxon 256 gene avian dataset by
\citet{Prum2015}.

\subsection{Evaluation procedure}
We study three questions in our evaluation:
\begin{itemize}
\item How accurate are branch lengths and support values 
when assumptions of our model are met?
\item How do violations of the model assumptions impact the result? 
\item How do local posterior probabilities compare to 
site-only MLBS?
\end{itemize}

To answer these questions, we use both true gene trees and estimated
gene trees to score both true and estimated species trees. 
For every internal branch in the species
tree being scored, we also score its two alternative topologies.
For example, if branch $AB|CD$ (Fig.~\ref{fig:basics})
appears in the species tree, we also score $AD|BC$ and $AC|BD$.
In our estimations, we use the
Yule prior with fixed $\lambda=\frac{1}{2}$, 
but note that the
true $\lambda$ in our A-200
simulations ranges from 
0.06 to 1.19. 

With true species trees and true gene trees, 
all model assumptions are met. 
When estimated
gene trees are used instead
of true gene trees,
we violate the assumption that input gene trees
follow properties of the MSC model.
When estimated species trees are scored,
the locality assumption is potentially violated (i.e., 
each of the four clusters around a branch 
may be incorrect).

\subsubsection{Measurement}

\paragraph{Posterior:}
Despite the long-standing debate
about correct interpretations of various measures of support~\citep[e.g.,][]{Salichos2013,Felsenstein1993,Hillis1993,Susko2009},
biologists typically use support
to judge branch reliability.
A common practice is to ignore branches
below a certain threshold of support and only interpret
the remaining branches as biologically meaningful
(0.95 for posterior and 70\% for bootstrap are often used). 
Our evaluation procedure takes a similar approach;
we use varying thresholds of support and 
count the number of true and false branches with 
support at least equal to the threshold. 
For a threshold $s$, the measures we use are 
precision (the percentage of branches with support 
$\ge s$ that are correct), recall (the percentage of all true branches
that have support $\ge s$),
and false positive rate (FPR) (the percentage of all false 
branches that have support $\ge s$).
We also draw the ROC curve (i.e., recall versus FPR). 

MLBS and
posteriors values are not directly comparable. 
Therefore, it is pointless to compare the precision
or recall of MLBS and posterior for a given threshold. 
Instead, 
we use the ROC curve, which is agnostic
to the exact interpretation of the threshold; it simply
shows which method results in a better trade-off
between false negative and false positive branches.  
Moreover, comparing to MLBS was only feasible on the avian
dataset, where gene bootstrapping was doable. 
On the A-200 dataset (300 replicates 
each with 1000 genes of 201 taxa) bootstrapping was not computationally
feasible.

\begin{table*}[!tb]
\tableparts{\caption{Precision (and recall) of local posterior probabilities on A-200 dataset\label{tab:lpp200}}}
{\tabcolsep=0pt\begin{tabular*}{\textwidth}{@{\extracolsep{\fill}}llcccccccc@{}}\toprule
&& \multicolumn{4}{c}{True species tree} &   \multicolumn{4}{c}{ASTRAL species  tree} \\
&& \multicolumn{2}{c}{True gene tree} &   \multicolumn{2}{c}{Estimated  gene tree}  &  \multicolumn{2}{c}{True gene tree} &   \multicolumn{2}{c}{Estimated  gene tree}  \\
&$\gns$&  .99  &  .95 &  .99  &  .95 &  .99 &.95 &  .99&.95  \\\colrule
Low ILS&1000&100.0(98.3)&100.0(98.7)& 98.6(94.6)& 98.4(95.4)& 98.8(98.4)& 98.8(98.8)& 98.0(95.1)& 97.8(95.9)\tabularnewline
Low ILS&200&100.0(95.9)&100.0(96.8)& 99.1(90.2)& 98.9(91.9)& 98.9(96.2)& 98.8(97.1)& 98.7(90.6)& 98.5(92.4)\tabularnewline
Low ILS&50&100.0(91.1)&100.0(93.2)& 99.6(81.0)& 99.3(85.0)& 98.7(91.6)& 98.6(93.6)& 99.4(81.6)& 99.0(85.6)\tabularnewline
Med ILS&1000&100.0(95.1)&100.0(96.2)& 98.9(90.8)& 98.7(92.4)& 99.3(95.4)& 99.2(96.6)& 98.6(91.2)& 98.4(92.8)\tabularnewline
Med ILS&200&100.0(89.2)&100.0(91.3)& 99.4(82.6)& 99.2(85.7)& 99.4(90.0)& 99.4(92.1)& 99.3(83.3)& 99.0(86.5)\tabularnewline
Med ILS&50&100.0(79.0)& 99.9(83.2)& 99.7(70.0)& 99.5(75.2)& 99.4(81.0)& 99.2(85.2)& 99.7(71.3)& 99.5(76.6)\tabularnewline
High ILS&1000&100.0(83.0)&100.0(86.0)& 99.4(77.5)& 99.2(81.3)& 99.7(84.2)& 99.6(87.1)& 99.4(78.3)& 99.2(82.1)\tabularnewline
High ILS&200&100.0(67.3)&100.0(72.8)& 99.8(60.3)& 99.6(66.6)& 99.8(69.6)& 99.7(75.1)& 99.8(61.9)& 99.6(68.3)\tabularnewline
High ILS&50&100.0(46.0)& 99.8(55.0)& 99.8(38.6)& 99.6(47.7)& 99.8(50.0)& 99.4(59.7)& 99.8(41.1)& 99.6(50.6)\tabularnewline
\botrule
\end{tabular*}}
{\tablenote{For local posterior probability thresholds
0.95 and 0.99, we show the precision and recall (shown parenthetically) when true or ASTRAL species 
trees are scored with true or estimated gene trees. }}
\end{table*}

\paragraph{Branch length:}
We measure branch length accuracy by comparing
 true and estimated lengths for each branch.
Since this can be done only for correct branches, 
we measure branch length accuracy
for the true species tree topology. 
Given $b$ branches, and letting
$w_i$ and  $\hat{w}_i$ indicate the true
and estimated branch lengths,
we use the logarithmic (log) error defined as $
\frac{1}{b}\sum_{1}^b  |\log_{10}(w_i) - \log_{10}(\hat{w}_i)|$.
We also plot log of estimated versus true
values. 
In addition, we show the root mean squared error, defined as $
\sqrt{ \frac{1}{b}\sum_{1}^b  (w_i - \hat{w}_i)^2|
}$.
On Avian datasets, 
we compare the error of MP-EST
and ASTRAL. 
\section{Results}


\subsection{A-200 dataset}

\subsubsection{Posterior}

\paragraph{True trees:}
When true species trees are scored with 
true gene trees, the precision 
of branches with 0.99 pp or higher 
is 100\% for all model conditions, and
is at least 99.8\% for the 0.95 threshold
(Table~\ref{tab:lpp200}).
Thus, there are very few false positive branches
that have high local pp, 
a trend that continues if we further lower
the threshold to 0.9 (Fig.~S2).
With the 0.95 threshold, the percentage of
true branches that are recovered (recall)
ranges from very high (98.7\%) for
the model condition with low ILS and 1000 gene trees
to moderate (55.0\%)
for the most challenging dataset with high ILS
and only 50 genes (Table~\ref{tab:lpp200}).
As desired, 
increasing ILS and reducing the number of genes
both reduce the recall while maintaining high precision (Fig.~S2).
 
 \newcommand{\cpr}[1]{
Evaluation of local posterior probability on the A-200 dataset
with #1 species trees.  See Figures S2-S4 for other species trees.
A) Precision and recall of 
branches with local posterior probability above a threshold
ranging from 0.9 to 1.0 using estimated gene trees
(solid) or true gene trees (dotted).
B) ROC curve (recall versus false positive rate) 
for varying thresholds (figure trimmed at 0.4 FPR). 
Columns show different levels of ILS}

\begin{figure*}[t]
\begin{center}
\includegraphics[width=0.75\textwidth]{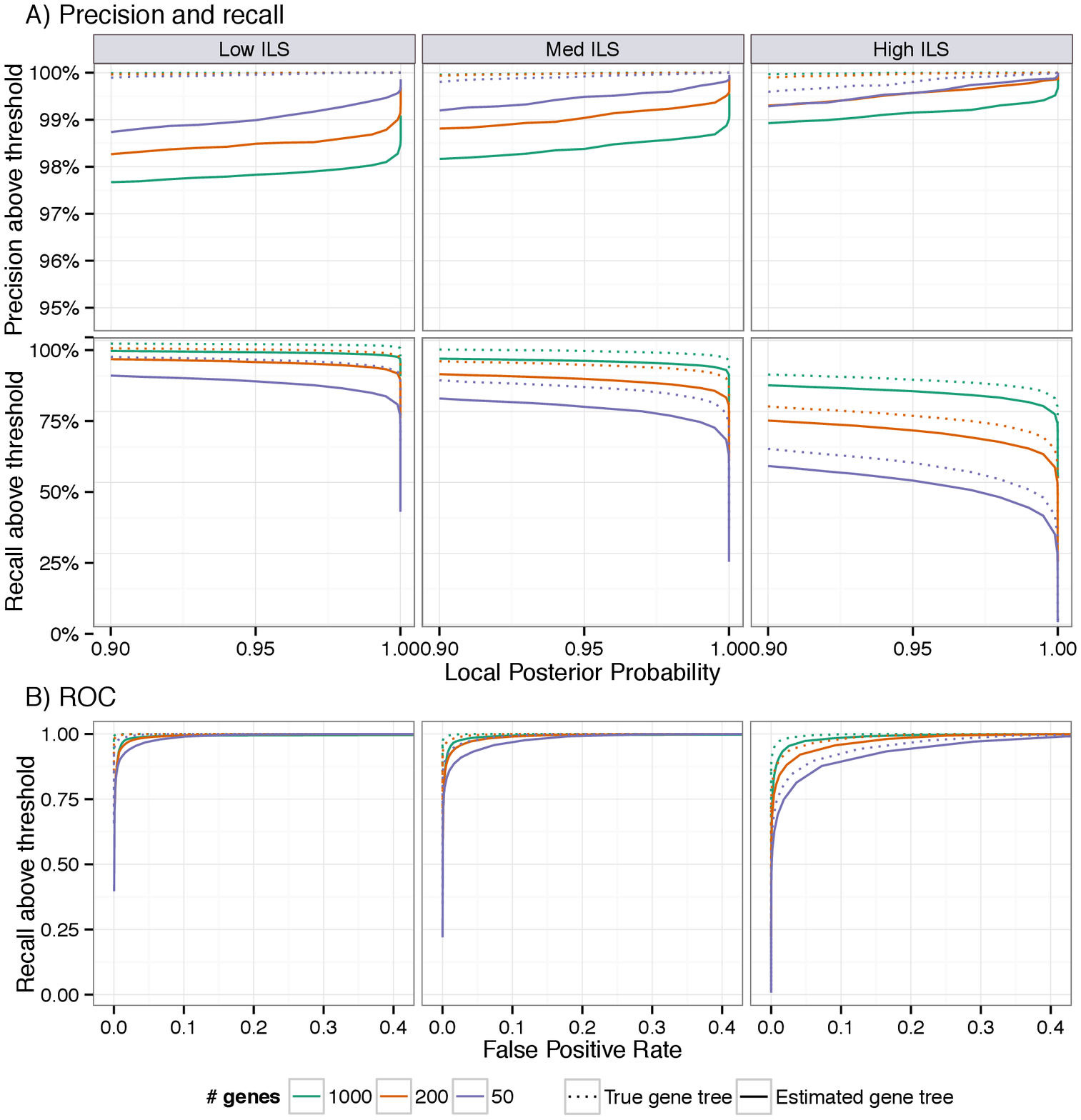}
\end{center}
\caption{\cpr{ASTRAL}.}%
\label{fig:200-pr}%
\end{figure*}

\paragraph{Estimated gene trees:}
When 
true 
species trees are scored on estimated
gene trees instead of true gene trees, 
precision slightly drops from 100\% to between 98.4\% and 99.8\%.
The recall, however, is impacted more and 
is reduced by as much as 10\% (Table~\ref{tab:lpp200} and
Fig. S2).
Thus, gene tree estimation error has a small
impact on the precision but a substantial impact
on the recall. 

The impact of the threshold is also interesting. 
Going from the 0.99 to 0.95 threshold, 
as expected, recall improves
 (e.g., from 39\% to 48\%
for high ILS, 50 genes)
but reductions in the precision are very small (at most 0.3\%).
Thus, a 95\% threshold results in meaningful
improvements in the recall without substantially
sacrificing the precision. 
The ROC curves (Figs.~\ref{fig:200-pr}B and S2) further
explore the tradeoff between
increasing recall and allowing false positive branches.

\paragraph{Estimated species trees:}
By scoring estimated species trees we 
study the impact of violating the locality
assumption. 
We show results for ASTRAL here, but 
similar results are obtained with NJst and concatenation (Figs. S2 and S4). 

Precision and recall are 
remarkably similar between ASTRAL and true
species trees, especially with estimated gene trees (Table~\ref{tab:lpp200}).
Comparing true species trees and ASTRAL
on estimated gene trees, 
the precision is reduced at most by 0.6\%
while the recall is surprisingly increased, by up to 2.9\%.
The impact of violating locality assumptions
is more pronounced when true gene trees are used. 
Once again, precision is reduced (by as much as
1.4\%) and the recall is increased (up to 4.7\%).
Thus, moderate violations of the
locality assumption have minimal impact.

We note that in our analyses, 
deviations from the locality
assumption are moderate but realistic, as most
ASTRAL trees have a relatively high accuracy
(Table~\ref{tab:datasets}). The
least accurate ASTRAL trees
have 19\% RF distance to the true species tree
(50 genes and high ILS), which means
81\% of the clusters in the estimated tree
remain correct. Nevertheless, 
it is interesting that violating the locality
assumption for up to 19\% of clusters
has minimal impact on the precision 
and positive impact on the recall.

 

\begin{table}[!tb]
\tableparts{\caption{Branch length accuracy on
the A-200 dataset. }\label{tab:BL}}
{\begin{tabular*}{\columnwidth}{@{\extracolsep{\fill}}ld{1}d{4}d{4}d{4}d{4}@{}}\toprule 
&&\multicolumn{2}{c}{Log Err} &\multicolumn{2}{c}{RMSE}  \\
Dataset&$\gns$&\mcc{True gt}&\mcc{Est. gt}&\mcc{True gt}&\mcc{Est. gt}\\
\colrule 
Low ILS&1000&0.10&0.42&5.57&6.75\tabularnewline
Low ILS&200&0.16&0.44&6.22&6.99\tabularnewline
Low ILS&50&0.25&0.48&6.84&7.29\tabularnewline
Med ILS&1000&0.03&0.20&0.22&0.86\tabularnewline
Med ILS&200&0.07&0.22&0.44&0.91\tabularnewline
Med ILS&50&0.13&0.26&0.74&1.05\tabularnewline
High ILS&1000&0.06&0.15&0.03&0.13\tabularnewline
High ILS&200&0.11&0.18&0.07&0.15\tabularnewline
High ILS&50&0.18&0.24&0.14&0.19\tabularnewline
\botrule
\end{tabular*}}
{Logarithmic error (Log Err)  and root
mean squared error (RMSE) are shown
for true species trees scored with true gene
trees or estimated gene trees (Est. gt). }
\end{table}

\begin{figure}[!htb]
\begin{center}
\includegraphics[width=0.5\textwidth]{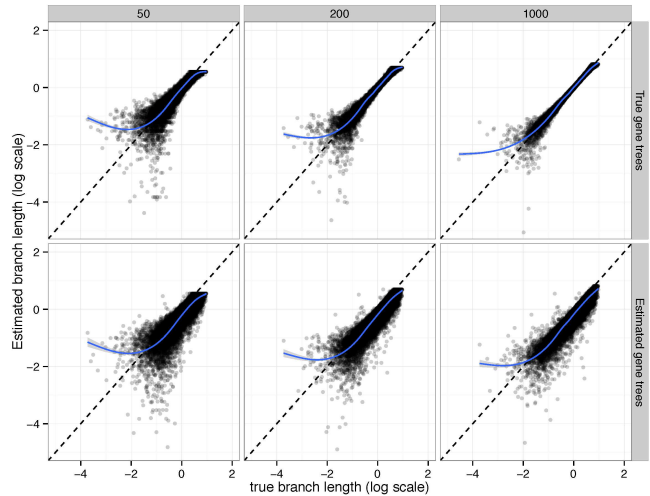}
\end{center}
\caption{Branch length accuracy on the A-200 dataset
with Medium ILS. See Figs S7 and S8 for low and
high ILS. The estimated branch length is plotted
against the true branch length in log scale (base 10).
Blue line: a fitted generalized additive model with smoothing~\citep{Wood2011}.}%
\label{fig:200-bl}%
\end{figure}

\subsubsection{Branch Length}
The accuracy of branch lengths is dramatically
impacted by gene tree estimation error, the number
of genes, and the amount of ILS (Table~\ref{tab:BL}). 
With 1000 true gene trees, 
the logarithmic error is very low, ranging
from 0.03 to 0.10 (which correspond to
branches that on average are respectively 7\% or 25\% shorter
or longer than true branches).
As the number of genes is reduced, the logarithmic 
error
predictably goes up, but with true gene trees, 
it never exceeds 0.25.
Moreover, with true gene trees,
the error is largely unbiased, except perhaps
for very short or long
branches that are hard to estimate correctly 
with a limited number of genes (Figs.~\ref{fig:200-bl} and S7-S8).

Branch length error dramatically increases
when estimated gene trees are used.
Low ILS conditions are impacted the most
by gene tree error (Table~\ref{tab:BL} and Fig.~S7).
For example, with 1000 estimated gene trees
and low ILS, log error is 0.42, corresponding
to estimated branches that are on average 2.6
times too short or long. 
Moreover, the error is
biased towards underestimation, especially
for low ILS
(Figs.~\ref{fig:200-bl}, S7, and S8).
This pattern is not surprising because
as we will show, gene tree error 
tends to increase observed
gene tree discordance and branch lengths
are a function of observed discordance.

\subsection{Avian}

On the avian dataset, we compare
local posterior probabilities against branch support
generated using site-only MLBS
with estimated gene trees and ASTRAL species trees.
Here, we also study the impact
of increasing levels of gene tree estimation error
by decreasing the number of sites per gene
from 1500bp to 250bp.

\subsection{Posterior and MLBS}

The precision of local posterior probability
is 100\% for the 0.99 threshold, regardless
of the numbers of sites, but the recall
ranges from 81\% for the 1500bp model condition
to 69\% for 250bp (Table~S1 and Figs. S5 and S6).
Precision is at least 99.8\% for 
the 0.95 threshold, and the recall 
is between 71.5\% and 84.7\%, depending on the model
condition (an improvement of 2\% to 5\%
compared to the 0.99 threshold). 
Lowering the support threshold all the way to 0.7 still 
retains at least 99.1\% accuracy
and increases the recall to between 78.3\% and 91.4\%. 
Therefore, the local posterior
probabilities
allow
very few false positives with high support but also
miss some true positives (and thus may
be conservative). 

\begin{figure}[tb]
\begin{center}
\includegraphics[width=0.49\textwidth]{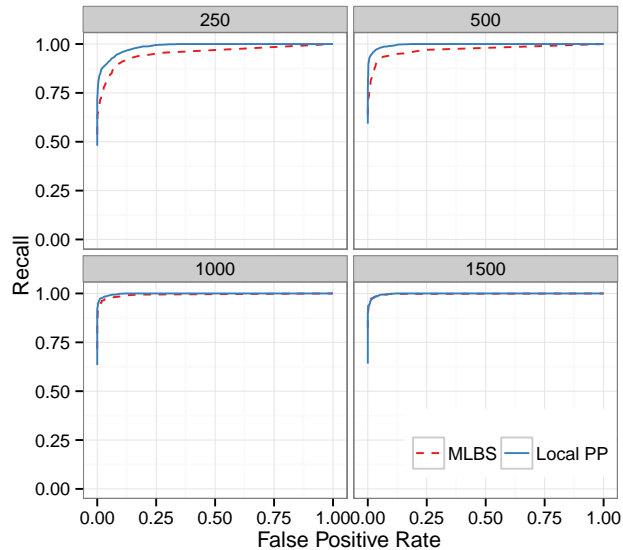}
\end{center}
\caption{ROC curve for the avian dataset based on 
MLBS and local posterior probability (PP)
support values.
Boxes show different numbers of 
sites per gene (controlling gene tree
estimation error). }%
\label{fig:roc-avian}%
\end{figure}

Nevertheless, local posterior probabilities
are less conservative 
than MLBS support values and have better recall.
As
the ROC curves
show (Fig.~\ref{fig:roc-avian}), for the same number of false positives branches,
local posterior probabilities result in better recall 
than MLBS.
This pattern is more pronounced for shorter
alignments, which have increased gene tree
error. 
For example, for the 250bp model condition, 
if we choose a support threshold that
results in 0.01 false positive rate, 
with local posterior values, we still recover 84\% of
correct branches, whereas with MLBS, 
the same false positive rate results in retaining
70\% of correct branches. 
Thus, for a desired level of precision, better
recall can be obtained using local posterior probabilities.

\begin{table}[!tb]
\tableparts{\caption{Branch length accuracy for the avian dataset. }\label{tab:BL-av}}
{\begin{tabular*}{\columnwidth}{@{\extracolsep{\fill}}llccc@{}} \toprule
& \multicolumn{2}{c}{Log Err} &   \multicolumn{2}{c}{RMSE}\\
\# sites&ASTRAL&MPEST&ASTRAL&MPEST\tabularnewline
\colrule 
True gt.&0.06 (0.10)&0.07 (0.11)&0.44 (0.44)&0.30 (0.30)\tabularnewline
1500&0.17 (0.20)&0.14 (0.18)&0.83 (0.83)&0.70 (0.70)\tabularnewline
1000&0.22 (0.27)&0.22 (0.25)&1.08 (1.07)&1.01 (1.00)\tabularnewline
500&0.37 (0.42)&0.42 (0.46)&1.65 (1.64)&1.65 (1.64)\tabularnewline
250&0.59 (0.63)&0.81 (0.84)&2.25 (2.24)&2.28 (2.26)\tabularnewline
\botrule
\end{tabular*}}
{Logarithmic error and root
mean squared error are shown
for true species trees scored with true gene
trees or estimated gene trees with various
numbers of sites using ASTRAL and MP-EST.
An extremely short branch with length $10^{-6}$ was
removed from the calculations, but
error including that branch is shown parenthetically.}
\end{table}

\subsubsection{Branch Length}

\begin{figure}[tb]
\begin{center}
\includegraphics[width=0.47\textwidth]{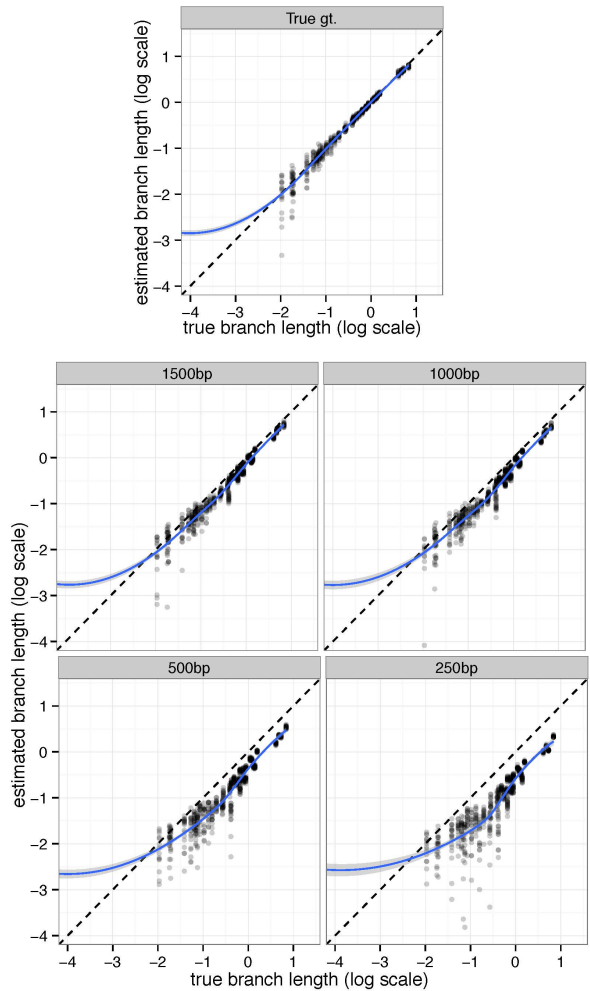}
\end{center}
\caption{ASTRAL branch length accuracy on the avian dataset.
Log transformed estimated branch lengths are shown
versus true branch lengths, and a generalized additive model 
is fitted to the data. One branch with length $10^{-6}$
is trimmed out here, but full results, including 
MP-EST, is shown in Fig. S9.}%
\label{fig:BL-avian}%
\end{figure}

Branch length accuracy on the avian dataset
was a function of gene tree
estimation error 
whether ASTRAL or MP-EST was used
(Table~\ref{tab:BL-av}). 
With true gene trees,
branch length log error was only 0.06, corresponding to branches
that are about 14\% shorter or longer than the true branch. 
As gene tree estimation error increases
with reduced number of sites (see Table~\ref{tab:datasets}
for gene tree error statistics), the branch
length error also increases. Thus, 
while 1500bp genes give 0.17 log error, 250bp genes
result in 0.59 error, which corresponds to branches that 
are on average 3.9 times too short or long. 
Moreover, unlike true gene trees, the error in 
branch lengths estimated
based on estimated gene trees is biased toward underestimation
(Fig.~\ref{fig:BL-avian}), a pattern
that increases in intensity with
shorter alignments. 

ASTRAL and MP-EST have similar
branch length accuracy measured by log
error for highly accurate gene trees, 
but ASTRAL has an advantage 
with increased gene tree error (Fig.~S10, 
Table~\ref{tab:BL-av}).
Error measured by RMSE
(which emphasizes the accuracy of long branches)
is comparable for 
the two methods, but
MP-EST has a slight advantage given accurate
gene trees. 

\subsection{Biological datasets}
For each biological dataset, we show
MLBS support and the local posterior probabilities, 
computed based on RAxML gene trees
available from respective publications. 
We also collapse gene tree branches with less than 
33\% bootstrap support and use these
collapsed gene trees to draw local posterior probabilities. 
For ease of discussion, we show local posterior probabilities
as percentages and refer to them simply as posterior
or collapsed posterior (for values based on collapsed gene
trees). 
We discuss the confidence in important branches in each tree.

\begin{figure*}[ht]
\begin{center}
\includegraphics[width=0.9\textwidth]{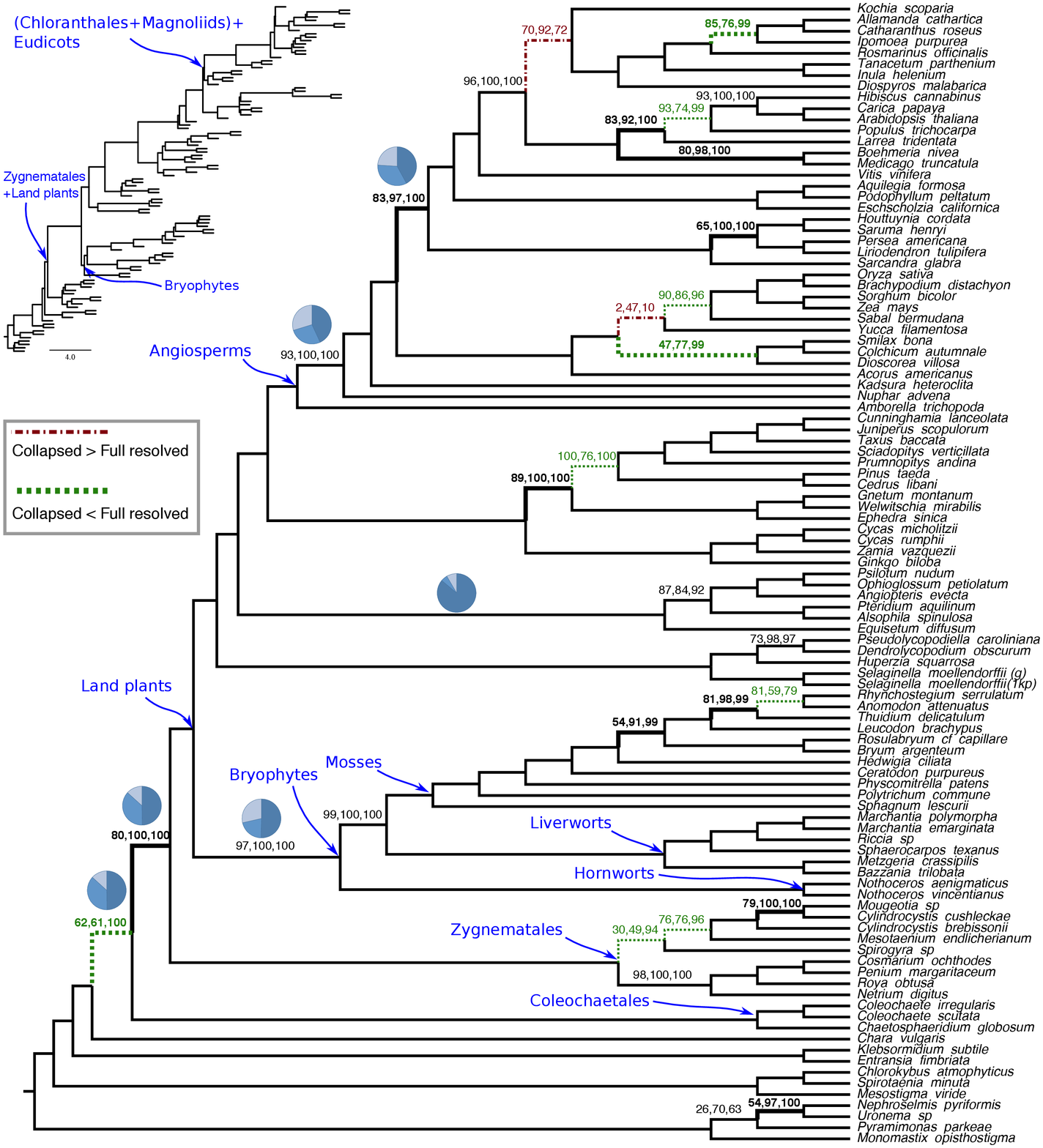}
\end{center}
\caption{ASTRAL tree on the 1KP
dataset of \citet{1kp-pilot} (103 taxa and 424 genes). On each branch,
three support values are shown: BS (using site-only MLBS), 
local posterior computed on fully resolved ML gene trees, 
and local posterior computed on collapsed ML gene trees 
(removing branches
with $<33\%$ BS). 
Branches with no designation have 100\% support according to
all three measures. 
Dotted/green lines (dashed/red lines): collapsing 
low support gene trees branches increases (decreases)
 posterior by at least 10\%. 
Bold: collapsed posterior is at least
10\% higher than BS.
Inset: ASTRAL tree with branch lengths in coalescent
units using collapsed genes (terminals lengths 
drawn arbitrarily).
Pie-charts (for selected edges): relative frequencies of the three
quartet topologies around a branch in 
collapsed gene trees. }%
\label{fig:1KP}%
\end{figure*}

\paragraph{1KP:}
Three of the key relationships studied
by \citet{1kp-pilot} are the sister branch
to land plants, the 
base of the angiosperms, and
the relationship among Bryophytes (hornworts, liverworts, 
and mosses).
In the ASTRAL tree, many branches
have full support regardless of the measure of the
support used, but the remaining branches
reveal interesting patterns (Fig.~\ref{fig:1KP}).
The sister relationship between Zygnematales
and land plants receives
a moderate 80\% BS, but
has 100\% posterior. 
\citet{1kp-pilot} also recovered 
this relationship by concatenation
of various data partitions. 
There are 12 other branches that 
have collapsed posteriors  that are
at least 10\% higher than BS (Fig.~\ref{fig:1KP});
no branch has substantially higher BS than  collapsed posterior.
Collapsed posterior for monophyly of Bryophytes
 and for Amborella as sister to other angiosperms are
100\% (compared to 97\% and 93\% BS, respectively). 

When we collapse low support branches in gene trees, posterior 
goes up for several branches:
nine branches have improvements of 10\% or more, and only
two branches have comparable reductions. 
An interesting case is
Coleochaetales as sister to Zygnematales+land plants, which
has only 62\% BS and 61\% posterior, 
but has 100\% collapsed posterior.  
Finally, note that several branches have low posterior,
even after collapsing. 

Our estimated branch lengths
are short for several nodes.
For example, the branch that unites
(Chloranthales+Magnoliids) and Eudicots 
has a length of 0.14 in coalescent units. 
Other branches
that have been historically hard to recover
also tend to have short branches; 
however, these are not necessarily 
extremely short branches that would implicate 
anomaly zone (two adjacent branches below 0.1
will result in an anomaly zone~\cite{Rosenberg2013}).
For example, Bryophytes had a length of 0.29, 
and  Zygnematales+land plants had a length of 0.28. 
These values, while short, are not below the often-cited
0.1 threshold. 
Moreover, as our simulation study showed, we caution
that branch lengths tend to be underestimated
because of gene tree error and these numbers should be treated
as lower bounds. 

\paragraph{Angiosperms:}
\citet{xi} have used 310 genes to study the base
of the angiosperm tree, 
a question of intense debate~\citep[e.g.,][]{Soltis2011,Zhang2012,Goremykin2013,Simmons2015}. Unlike the MP-EST tree by~\citet{xi}, but 
similar to concatenation on this dataset and 
ASTRAL and concatenation on 1KP, 
the ASTRAL tree 
recovers Amborella as sister to the rest of the angiosperms.
This relationship has 75\% BS, 
but its posterior and collapsed posterior 
are 100\% (Fig. S11). 
The length of this branch is estimated
to be 0.160, almost exactly
matching the length estimated on the
1KP dataset (0.156).

\paragraph{Avian (genomes):}
\citet{avian} used whole-genomes
of 48 bird species to resolve
long-standing questions about relationships at
the base of Neoaves. 
We reanalyzed their 2022 supergene trees (binned gene trees;
see~\cite{binning}) using ASTRAL,
which produced a tree
with a wall of short branches at the base
of Neoaves (Fig. S12); 12 branches 
are
below 0.1 coalescent units, and another 11 are
below 0.5. 
However, when low support branches in gene trees
were collapsed, branch lengths increase by a median
of 0.23 units. Nevertheless, 11 branches
remain below 0.1, and another
four branches are below 0.5. 
Our results support the hypothesis that 
avian big bang~\cite{Feduccia2003} gave rise to
very short branches, but quantifying the exact lengths 
remains difficult because of gene tree error. 

Support values on the avian tree also revealed interesting
patterns. Despite the large number
of supergene trees (2022), which should increase
support (Fig.~\ref{fig:example}), 
several key branches
have low posterior. 
For example, the position of Hoatzin
(arguably, the most difficult avian order to place)
and the two branches
around it have collapsed posterior 
below 50\%  and posterior below 75\% (Fig. S12).
Similarly, at the base of Neoaves, a clade containing
land birds, water birds, and Caprimulgiformes 
has full support, but the sister to this large group
has only 61\%
posterior and 80\% collapsed posterior. 
However, other challenging relationships have full support
(e.g., falcons as sister to parrots+passerines, and seriema as
sister to this group, or a clade containing owls, eagles, and vultures). 
Thus, despite large number of input trees, 
posterior probabilities reveal some uncertainty. 
Finally, 
on this dataset, unlike the 1KP dataset, four branches
have substantially lower posterior
compared to BS, and posteriors 
are higher than collapsed posteriors in some cases. 

\paragraph{Avian (high sampling):}
\citet{Prum2015} used their dataset of 259 genes
and 201 species 
to study the avian tree with high
taxon sampling. The
ASTRAL tree reported by \citet{Prum2015} 
has low MLBS (Fig. S13), and
 many branches
remain poorly supported with posterior probabilities (the median difference
between BS and collapsed posterior was 0).
Moreover, 
many of the most interesting relationships are poorly supported. 
For example, the sister to parrots+passerine has
only 30\% MLBS support, 83\% local posterior support, 
and  0\% collapsed posterior.
These low support values are encouraging
because
the sister to parrots+passerine is likely recovered
incorrectly in this tree, as most
recent studies put falcons as sister to this group~\cite{avian,Suh2011,Kimball2013,McCormack2013}.
Overall, despite its large taxon sampling, this
dataset provides little resolution for the 
early Neoaves radiation using ASTRAL because
of the insufficient gene count for this high level of ILS.  
Just like the avian genomic data, here we obtain a wall
of short branches around the assumed rapid radiation 
of Neoaves (Fig S13). 

\section{Discussions}

The local posterior probabilities introduced
in this paper can be computed
quickly and without a need 
for extensive MCMC sampling
or bootstrapping. 
Computing posteriors for a species
tree with 200 taxa and 1000 genes
takes only 10 seconds and for a 
dataset with 1000 taxa and 1000 genes, 
takes about three minutes on a laptop
machine. 
This extremely fast computation is possible
only because of the two main assumptions
of the method: that true MSC-generated
gene trees
are given and the locality assumption
(i.e., the four clusters around
each internal branch are present in the true 
tree).
These assumptions can both be violated on real data.
Recognizing this fact, our simulations include
conditions that violate these two assumptions 
by introducing plenty of gene tree
estimation error (ranging from average
RF distance of 25\% to 67\%) 
as well as species tree error
(Table~\ref{tab:datasets} and Fig. S1). 

Our method 
allows very few false positives with 
high support, a
pattern that is retained even
with high levels of gene tree estimation error. 
It could be argued that our method is perhaps too conservative 
and underestimates support.
Nevertheless, local posterior probabilities
were {\em less} conservative than MLBS, the only viable alternative for large
datasets.
Despite allowing very few
false positives with high support,
the method generally had high recall
(i.e., true branches with high support)
except for very few genes for a
given amount of ILS. 
Reassuringly, increased gene tree estimation error
only negatively impacted recall
but retained very high precision. 
While underestimation of support is not desirable, 
the abundance of 
false branches with high support
would be a more serious problem.

A practical question is at what threshold of 
support a branch can be judged 
reliable. The answer depends on 
factors such as the false positive rate
desired and  the amount of gene
tree error. Nevertheless, it seems that 
the commonly used 0.95 threshold 
results in very high precision while retaining moderately high
recall. In our analyses, even lower thresholds
(e.g., 0.9 or even 0.7) give high precision, while
increasing the recall. 

\subsection{Gene tree estimation error}

An interesting pattern
was that with estimated gene trees (but not with true 
gene trees), at a given threshold, 
support values are more precise for high 
ILS compared to low ILS (Fig.~\ref{fig:200-pr}A).
We postulate this effect is related 
to the larger impact that gene tree estimation error
has on the total amount of observed discordance
for low ILS compared to high ILS conditions. 
Consistent with this explanation, we also 
observed a larger degradation of 
branch length accuracy in going
from true to estimated gene trees 
for low ILS conditions compared to high ILS
(Table~\ref{tab:BL}).

\begin{figure}[t]
\begin{center}
\includegraphics[width=0.5\textwidth]{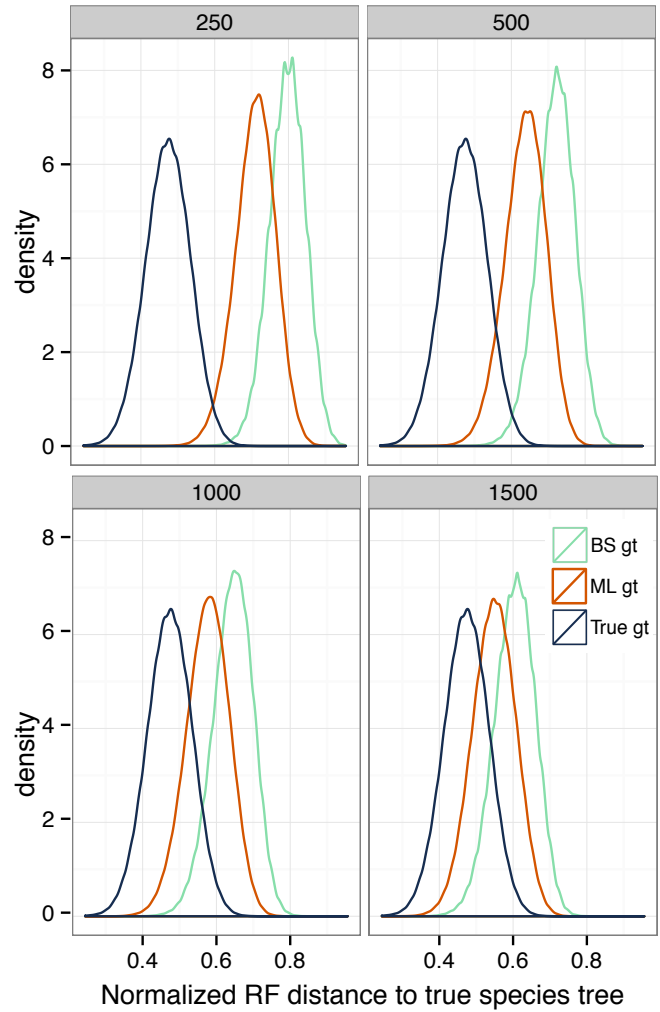}
\end{center}
\caption{Gene tree discordance for the avian dataset. 
We show the density plot of the normalized
RF distance between the true species tree and true gene trees, 
ML gene trees, and BS gene trees for four different model conditions. }%
\label{fig:disc}%
\end{figure}

The issue of gene tree estimation error is at the heart
of why we saw a need for developing this new
method. 
Sets of site-resampled bootstrap gene trees 
tend to have
increased levels of discordance
with regard to the species tree (and also among themselves)
compared to ML gene trees,
especially when each gene has a limited phylogenetic
signal. Bootstrapped gene trees have much higher rates of discordance
than either true gene trees or ML gene trees (Fig.~\ref{fig:disc}). 
It is expected that bootstrapped replicates of a dataset result in 
noisier estimates of parameters than ML; however,
the added error by bootstrapping should not be {\em biased}.
For MLBS, the input to the summary method is
not just a noisy dataset, but a {\em biased} one with 
increased levels of discordance.
We postulate this bias is the reason for the underperformance
of MLBS. 

Our method, on the other hand, does
not require bootstrapping and uses the best
available gene trees (e.g., estimated ML gene trees). 
While ML gene trees are still biased towards increased
discordance (and hence reduced branch length), 
they are better than bootstrapped gene trees (Fig~\ref{fig:disc}). 
The downside of our approach is that
gene tree uncertainty is not considered directly.
Thus, it was reassuring to see that our method
remains precise with high gene tree estimation 
error. 
To account for gene tree
estimation error, one can collapse the most
poorly resolved branches in gene trees
when computing support.
As we show in our analyses of
biological data, this practice seems promising. 
However, we note that when collapsing
branches, one should be careful not to introduce
bias, which can happen with aggressive 
filtering. We choose to 
collapse branches with support below 33\% (which 
can be considered randomly resolved). 
Future work needs to further study the effect
of collapsing low support branches in gene 
trees on both branch length and support. 

The impact of gene tree estimation error was
most clear with estimates of the branch length.
The branch lengths produced
by our method showed encouraging 
patterns (e.g., consistency across biological
datasets);
nevertheless, our estimated 
branch lengths are not immune to underestimation
that is seen often with other summary methods. 
Thus, we suggest branch lengths
from ASTRAL and other summary methods
should be interpreted with care in the presence
of gene tree estimation error. 

\subsection{High support despite high discordance}

An important observation, predicted
by the theory but sometimes lost in
the scientific debate about discordance, 
is that high confidence for a correctly 
inferred relationship can emerge
even with high levels of discordance. 
As Figure~\ref{fig:example} shows, 
a branch that appears in only 40\%
of gene trees can still be resolved
with high confidence
if a sufficient number of genes are available
(e.g., around 500). 
For example, in the 1KP tree, 
the branch that puts Zygnematales
as sister to land plants appeared in only 
49\% of collapsed gene tree quartets
and the branch making Bryophytes
monophyletic only appeared in
50\% of them; both 
branches, however, have a posterior of 1.0. 
We have implemented an option in
ASTRAL to output the percentage of 
gene tree quartets that agree with each 
of the three resolutions around a branch. 
Pie charts in Figure~\ref{fig:1KP}
give examples of these relative
quartet frequencies.

A question that biologists often face
is the number of genes required to resolve a 
branch. The number of genes required to obtain
high resolution and low false positive rates depends
on the model condition. With higher ILS, 
more genes are required, an observation
that is not surprising.
However, our method can be extended 
to estimate the number of genes that might be
required to resolve a tree (with an estimated level of ILS). 

\subsection{Limitations and future work}

Promising approaches for incorporating gene tree
uncertainty into local posterior probabilities exist. 
For example, one can weight each gene tree quartet around a 
branch by its SH-like
support, BS, or advanced
measures like concordance~\cite{Ane2007}.
Moreover, 
comparing our method against Bayesian co-estimation methods on 
small datasets where they can run will  
be interesting. 
Furthermore, we did not investigate the impact of changing
prior parameters ($\lambda$); nor did we explore other 
prior functions, such as Dirichlet distributions (conjugate to multinomial)
or birth-death processes. 
We leave these  for future work.  

We violated
some but not all assumptions of our method in the 
experimental results. 
The sequence evolution models used for simulation
and inference were both GTR+$\Gamma$, 
but on real data,  model violations
(e.g., compositional bias) can lead to biased
estimates of gene trees. 
Finally,
all our simulated datasets had discordance that was generated
only by ILS and estimation error, and not
other sources of true biological discordance, such 
as undetected paralogy or horizontal 
gene transfer.
Future work should further examine the
impact of other biological
sources of discordance on the reliability
of local posterior probabilities. 

\section{Supplementary Material}
Supplementary table S1– and figures S1–-S15 
are available  at Molecular Biology and Evolution
online (http://www.mbe.oxfordjournals.org/).

\section{Acknowledgments}
The authors gratefully thank
Prof. Tandy Warnow and Prof. Jim Leebens-mack
for helpful comments on earlier versions of this manuscript. 
For computational resources, this work
heavily relied on the Gordon cluster made generously available to us
on the San Diego Supercomputer Center (SDSC) cluster and 
through the Extreme Science and Engineering Discovery Environment (XSEDE), 
which is supported by National Science Foundation grant number ACI-1053575.

 \clearpage

\bibliographystyle{natbib}
\bibliography{localpp}

\clearpage
\end{document}